\documentclass[aps,prx,twocolumn,10pt]{revtex4-2}
\usepackage[utf8]{inputenc}
\usepackage{bm}
\usepackage{bbm}
\usepackage{bbold}
\usepackage{amsfonts}
\usepackage{amsmath}
\usepackage{amssymb}
\usepackage{graphicx}
\usepackage{mathtools}
\usepackage{upgreek}
\usepackage{complexity}
\usepackage{xfrac}
\usepackage{soul}
\usepackage{xcolor}
\usepackage{dsfont}

\usepackage[colorlinks=true,citecolor=blue]{hyperref}
\hypersetup{colorlinks=true,citecolor=blue,linkcolor=blue,urlcolor=blue}
\usepackage{url}

\usepackage{txfonts}
\DeclareSymbolFont{matha}{OML}{txmi}{m}{it}
\DeclareMathSymbol{\varv}{\mathord}{matha}{118}

\newcommand{\defeq}{\vcentcolon=}
\newcommand{\eqdef}{=\vcentcolon}
\newcommand{\slashedzero}{\mathrlap{0}\kern0.25em\mathchar"323}

\begin{document}

\title{Can Geometric Quantum Machine Learning Lead to Advantage in Barcode Classification?}

\author{Chukwudubem Umeano}
\affiliation{Department of Physics and Astronomy, University of Exeter, Stocker Road, Exeter EX4 4QL, United Kingdom}

\author{Stefano Scali}
\affiliation{Department of Physics and Astronomy, University of Exeter, Stocker Road, Exeter EX4 4QL, United Kingdom}

\author{Oleksandr Kyriienko}
\affiliation{Department of Physics and Astronomy, University of Exeter, Stocker Road, Exeter EX4 4QL, United Kingdom}

\date{\today}

\begin{abstract}
We consider the problem of distinguishing two vectors (visualized as images or barcodes) and learning if they are related to one another. For this, we develop a geometric quantum machine learning (GQML) approach with embedded symmetries that allows for the classification of similar and dissimilar pairs based on global correlations, and enables generalization from just a few samples. Unlike GQML algorithms developed to date, we propose to focus on symmetry-aware measurement adaptation that outperforms unitary parametrizations. We compare GQML for similarity testing against classical deep neural networks and convolutional neural networks with Siamese architectures. We show that quantum networks largely outperform their classical counterparts.
We explain this difference in performance by analyzing correlated distributions used for composing our dataset.
We relate the similarity testing with problems that showcase a proven maximal separation between the $\BQP$ complexity class and the polynomial hierarchy. While the ability to achieve advantage largely depends on how data are loaded, we discuss how similar problems can benefit from quantum machine learning.
\end{abstract}

\maketitle


\textit{Introduction.---} Machine learning (ML) became a crucial tool for information processing \cite{LeCun2015}, bringing us closer to advanced artificial intelligence \cite{moskvichev2023concept,Biever2023-kv}. Its further progress largely depends on the development of qualitatively new hardware for computing and algorithmic breakthroughs \cite{Bakir2007}. From this perspective, quantum machine learning (QML) \cite{Benedetti2019rev} --- an area of ML devoted to developing quantum computing algorithms to solve learning-based problems --- has attracted much attention in recent years \cite{Biamonte2017,Havlicek2019,Liu2024}. Generally motivated by the capability of quantum computers of getting a superpolynomial improvement in scaling as compared to classical algorithms \cite{Shor1997,nielsen2010quantum,Dalzell2023rev}, QML has evolved both in terms of applications to be considered, and approaches to be used. Notably, the adoption of variational quantum algorithms (VQAs) \cite{Cerezo2021rev,Tilly_2022} led to the variety of quantum circuit optimization-based protocols, applied to tasks such as classification \cite{mitarai2018quantum,PerezSalinas2020datareuploading,Havlicek2019,Huang2021NatComm,Nghiem2021,Li2021rev}, phase recognition \cite{Cong2019,Monaco2023,Herman2023}, generative modeling \cite{Liu2018,Zoufal2019,Coyle2020,Paine2021,kyriienko2022protocols,kasture2022protocols,Hibat-Allah2024}, and physics-informed solving of differential equations \cite{kyriienko2021solving,lubasch2020variational,Markidis2022qpinns,heim2021quantum,paine2023quantum,paine2023physics}, showing potential in areas where classical methods struggle. A distinct set of tools has been developed for embedding data into quantum computers \cite{Schuld2019feature,Abbas2021,Goto2021PRL}, including various feature maps designed to exploit the high-dimensional space of quantum states for enhanced data representation and processing efficiency \cite{Altares-Lopez_2021,williams2023quantum,Albrecht2023,jaderberg2023let,Umeano2024groundstate,Wu2024Hartley,thabet2024quantumpositional}. The use-cases of QML include image recognition \cite{Chang2022,Schetakis2022,Senokosov_2024,shen2024classificationfashion}, data analysis in high-energy physics \cite{Belis2021,SChen2022,wozniak2023quantum,Belis2024,DiMeglio2024}, financial risk assessment \cite{Egger2020,Herman2023,Leclerc2023}, fraud detection \cite{kyriienko2022unsupervised,Grossi2022fraud,doosti2024briefreview,Innan2024}, natural language processing \cite{coecke2020foundations,Widdows2024nearterm,widdows2024QNLPrev}, graph-based learning \cite{verdon2019quantumgraph,LPHenry2021,dalyac2024graphalg,yu2023quantumgraph}, power flow analysis \cite{Kaseb2024}, enhanced molecular simulations in chemistry \cite{Kiss_2022}, among many others.

Several open questions are yet to be answered before QML delivers its full promise. First, we need to decide how to parameterize quantum models in the most advantageous way. Do we get an advantage from training variational circuits, or is quantum embedding enough \cite{Cerezo2023CSIM}? Or shall we use the structure of quantum kernels \cite{schuld2021supervised,gyurik2023limitations,Jerbi2024}? Second, we still need to understand where a proven advantage in quantum machine learning can come from \cite{Schuld2022PRXQ}. Does it mainly concern improving runtime and circuit depth \cite{Wiebe2012,lloyd2013quantum,Biamonte2017}? Does it come from sampling (generative modelling) \cite{Mansky2024,Hangleiter2023,kyriienko2022protocols,Gili_2023}, or do we gain from better generalization of quantum models \cite{Huang2022,Caro2022,Cong2019,umeano2023learn,gilfuster2023understanding}?
\begin{figure}[b!]
\includegraphics[width=1.0\linewidth]{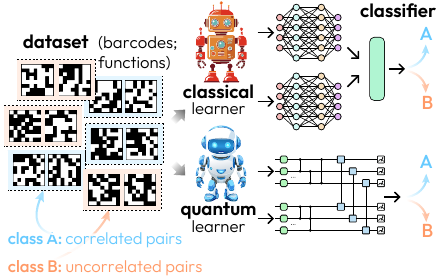}
    \caption{Sketch of the machine learning workflows for distinguishing images (barcodes) based on their correlations. Pairs of barcodes are processed by a classical learner via Siamese deep neural networks. A quantum learner processes data embedded into quantum states via parametrized equivariant measurements, classifying barcodes as correlated or uncorrelated.}
    \label{fig:sketch}
\end{figure}

One idea fuelling efficient quantum models leans on using symmetries of considered datasets and problems \cite{ZhengStrelchuk2023,JJMeyer2023PRXQ,Larocca2022PRXQ}. This approach, inspired by geometric deep learning \cite{bronstein2021geometric}, was dubbed as geometric quantum machine learning (GQML) \cite{nguyen2022theory,ragone2023representation,Cenk2024,le2023symmetryinv}. GQML excels in problems with apparent symmetries and offers a guide for optimal ansatz parametrizations. But we need to keep in mind that symmetries also help classical ML making problems more tractable \cite{Anschuetz2023efficient,bespalova2021quantum}.

To answer the second question several approaches were suggested. One can relate QML with probably approximately correct (PAC) learning \cite{Valiant1984}, proving a superpolynomial advantage for density and Boolean function learning \cite{Sweke2021quantumversus,Pirnay2023}, and for specific concept classes \cite{gyurik2023limitations,gyurik2023expsepar}. Another approach to proving quantum-classical separation relies on connecting QML models to some of the known quantum algorithms that have exponential separation. This includes developing quantum kernels with provable advantage based on $\BQP$ problems \cite{Liu2021NatPhys,Jager2023} and kernel-based quantum models with Grover's advantage \cite{muser2023provable}. Recently, we have shown that geometric quantum machine learning can be used to learn Simon's algorithm in $\BQP^A$ \cite{umeano2024geometric}, suggesting that both QML architectures and optimal circuits can be learnt to represent known solutions with proven advantages. 

In this paper, we tell a story of GQML being applied to the similarity testing problem. We show that for data sampled from specific distributions, certain classes of GQML models can provably outperform classical learners in terms of generalization, discovering an optimal decision boundary with just few samples. We set up the problem as distinguishing pairs of pixelated images---barcodes---each represented by a binary string. The task is to understand which pairs are related to each other (i.e., coming from a correlated distribution), or are dissimilar. We develop several architectures for an equivariant quantum neural network (QNN) to address the task. We show that, while the unitarily-parametrized QNN does not train well on the supplied dataset, the adaptive measurement-based QNN trains and provides a \emph{generalization} advantage in learning. We explain the results based on established problems for superpolynomial separation between $\BQP^A$ and $\PH$.


\textit{The problem.---}We consider a dataset $\mathcal{D}$ consisting of samples labelled as correlated ($y_m = 0$, class A) and uncorrelated ($y_m = 1$, class B). The dataset $\mathcal{D} = \{ x_m = (\mathbf{x}_{1,m},\mathbf{x}_{2,m}), y_m \}_{m=1}^{M}$ of $M$ samples includes information about $x_m$ being pairs of binary strings $\mathbf{x}_{1,m} \in \mathcal{X}$ and $\mathbf{x}_{2,m} \in \mathcal{X}$, where $\mathcal{X} = \{0,1\}^N$. The first part of the pair (e.g. $\mathbf{x}_{1,1} = 010001...$) and the second part ($\mathbf{x}_{2,1} = 101100...$) need to be loaded and processed in parallel. Classically, these samples can be represented as two-dimensional barcodes with filled and empty squares. The task is to establish a machine learning workflow for the similarity testing between the pairs (Fig.~\ref{fig:sketch}). We are given the information that samples come from latent distributions as $x_{|y=0} \sim P_{\mathrm{A}}(x)$ and $x_{|y=1} \sim P_{\mathrm{B}}(x)$, where $x \in \mathcal{X} \times \mathcal{X} \eqdef \mathcal{X}^{2}$. Therefore, the underlying goal is to distinguish the probability distributions $P_{\mathrm{A}}(x)$ and $P_{\mathrm{B}}(x)$ in terms of their properties related to \emph{global} correlations between samples coming in pairs. 

We suggest that the data can be efficiently encoded into quantum states using the phase states $|\phi_{x}\rangle  \defeq \Big(\sum_{j=1}^{N} (-1)^{\mathbf{x}_1^{[j]}}|j\rangle\Big) \otimes \Big(\sum_{j=1}^{N} (-1)^{\mathbf{x}_2^{[j]}}|j\rangle\Big)/N \eqdef |\phi_{\mathbf{x}_1}\rangle \otimes |\phi_{\mathbf{x}_2}\rangle$, where $\mathbf{x}_\alpha^{[j]}$ ($\alpha = 1,2$) denote the $j$-th digit of the binary converted into a $0/\pi$ phase in front of the corresponding computational basis state \cite{nielsen2010quantum}. In this case, $N$-digit strings are compressed into $n = \lceil \log_2 N \rceil$ qubit registers, and we need $2n$ qubits for loading a sample $x_m$. We are also given an associated symmetry $\mathcal{S}_\Phi$ that corresponds to a remapping of bitstrings according to rules given by particular Boolean circuits, such that $x_k$ and $\sigma(x_k)$ for $\sigma \in \mathcal{S}_\Phi$ share the same label $y_k$. This has the representation of $\sigma: x_k \mapsto \sigma(x_k) \cong \bar{x}_k$, where $\bar{x}_k = (\bar{\mathbf{x}}_{1,k},\bar{\mathbf{x}}_{2,k})$ is the complement of $x_k$, crucially containing the same global correlations. At this point, we are ready to propose machine learning approaches that can address the problem. 
\begin{figure}[t!]
\includegraphics[width=0.9\linewidth]{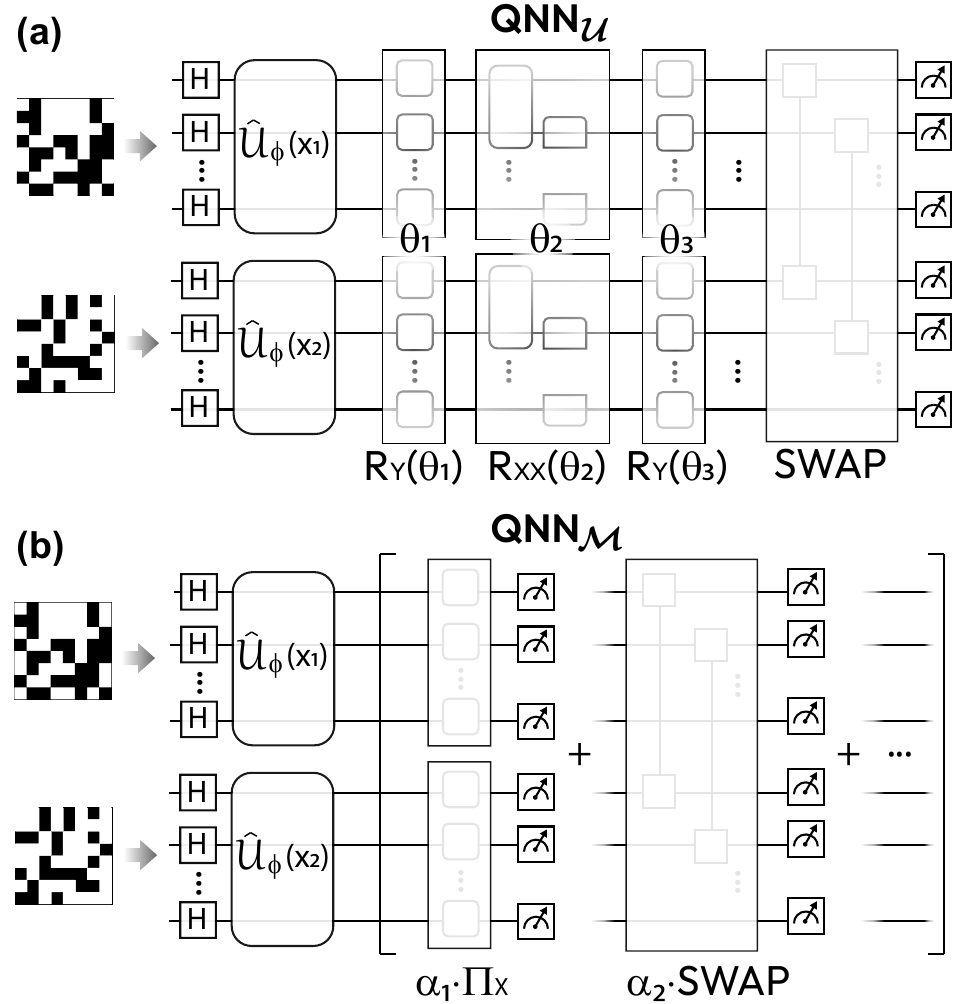}
    \caption{Architectures for geometric quantum machine learning used in the study. \textbf{(a)} Quantum neural network approach based on equivariant unitary circuit (QNN$_{\mathcal{U}}$) parametrized by a vector $\bm{\theta}$ that processes a pair of barcodes encoded into quantum states. \textbf{(b)} QNN$_{\mathcal{M}}$ based on equivariant measurements that are weighted adaptively to form a model for correlated sample processing.}
    \label{fig:architectures}
\end{figure}

\textit{The architectures.---}Next, we detail distinct machine learning approaches and neural network architecture that can be used to solve the problem of similarity testing. Specifically, we describe several QNNs and loss functions originating from geometric quantum machine learning. Then, we describe the classical workflows to set a baseline for benchmarking. 

For the GQML, we consider two different options. The first is the standard formalism for GQML that has been detailed in various studies \cite{JJMeyer2023PRXQ,nguyen2022theory,Larocca2022PRXQ,ZhengStrelchuk2023,bowles2023contextuality}. We wish to determine whether each pair of barcodes $x_m$ drawn from the dataset $\mathcal{D}$ is correlated or not. In other words, we have a target function $g: \mathcal{X} \times \mathcal{X} \rightarrow \mathcal{Y}$ mapping samples to the binary domain such that $g(x_m)=y_m$, and we wish to train a model to approximate $g$ across all inputs $x_m$. The quantum model is defined as
\begin{equation}
    h_\theta (x_m)=\langle\psi_0|\hat{\mathcal{U}}_{\phi}^\dagger(x_m)\hat{W}^\dagger(\theta)\hat{\mathcal{O}}\hat{W}(\theta)\hat{\mathcal{U}}_{\phi}(x_m)|\psi_0\rangle,
\end{equation}
with $|\psi_0\rangle = H^{\otimes 2n}|0\rangle^{2n}$ being a uniform state and $\hat{\mathcal{U}}_{\phi}(x)$ being the feature map that creates a phase state as $|\phi_x \rangle \eqdef \hat{\mathcal{U}}_{\phi}(x) |\psi_0\rangle$. Next, the state is processed with an adjustable ansatz $\hat{W}(\theta)$ followed by the measurement of a Hermitian observable $\hat{\mathcal{O}}$. 

To move from QML to GQML, we choose the components of our model in a way that incorporates symmetries and ensures label invariance. Given a symmetry group $\mathcal{S}$ with a corresponding representation $V_\sigma: \mathcal{S} \rightarrow Aut(\mathcal{X}^{2})$ \cite{JJMeyer2023PRXQ,nguyen2022theory}, we say that the function $g$ is invariant under the action of $\mathcal{S}$ if $g(V_\sigma[x_m])=g(x_m)$ $\forall x_m \in \mathcal{X}^{2}, \sigma \in \mathcal{S}$. In our case, we have the symmetry group $\mathcal{S} = \mathds{Z}_2 \times \mathcal{S}_\Phi$. The first symmetry corresponds to the exchange of two barcodes within the pair, $g(V_{Z_2}[\mathbf{x}_{1,m},\mathbf{x}_{2,m}]) = g(\mathbf{x}_{2,m},\mathbf{x}_{1,m}) = g(\mathbf{x}_{1,m},\mathbf{x}_{2,m})$, while the second corresponds to the bitstring remapping symmetry applied on each barcode, as described in the previous section: $g(V_{S_\Phi}[\mathbf{x}_{1,m},\mathbf{x}_{2,m}]) = g(\bar{\mathbf{x}}_{1,m},\bar{\mathbf{x}}_{2,m}) = g(\mathbf{x}_{1,m},\mathbf{x}_{2,m})$.

To encode these symmetries on the Hilbert space, we use the unitary representations of our symmetries: $\hat{U}_{Z_2} = \prod_{i=1}^n \mathrm{SWAP}_{i,i+n}$ for the exchange symmetry, and $\hat{U}_\Phi = Y^{\otimes n} \otimes Y^{\otimes n}$ for the bitstring remapping symmetry. Guided by these, we can build an invariant quantum model using the following ingredients \cite{JJMeyer2023PRXQ}: 1) \emph{invariant} initial state, $\hat{U}_\sigma|\psi_0\rangle = |\psi_0\rangle$; 2) \emph{equivariant} embedding, $\hat{U}(V_\sigma[x_m])=\hat{U}_\sigma\hat{U}(x_m)\hat{U}_\sigma^\dagger$; 3) \emph{equivariant} ansatz, $[\hat{W}(\theta), \hat{U}_\sigma] = 0$; and 4) \emph{invariant} measured observable satisfying $\hat{U}_\sigma\hat{\mathcal{O}}\hat{U}_\sigma^\dagger = \hat{\mathcal{O}}$.
\begin{figure}[t!]
    \centering
    \includegraphics[width=0.95\linewidth]{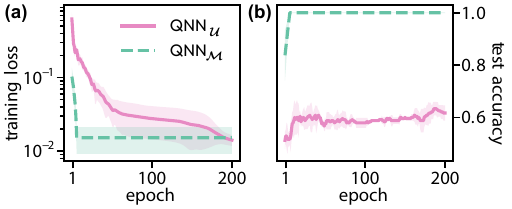}
    \caption{Comparison of the training performance \textbf{(a)} and generalization \textbf{(b)} between the two GQML architectures, $\text{QNN}_\mathcal{U}$ and $\text{QNN}_\mathcal{M}$. Both models were trained and tested on pairs of barcodes with 16 pixels each, corresponding to an 8-qubit quantum state for each input datum. Solid/dashed lines indicate the mean values across the 10 trials, while the shaded region indicates the standard deviation.}
    \label{fig:measVunitary}
\end{figure}

The equivariant embedding of the pairs of barcodes is done by preparing the aforementioned phase states $|\phi_{x_m}\rangle$. 
These are also known as real equally weighted (REW) states \cite{qu2012quantum} or hypergraph states \cite{Rossi_2013}, and the corresponding unitary encoding circuit $\hat{\mathcal{U}}_{\phi}(x_m)=\hat{U}(\mathbf{x}_{1,m})\otimes\hat{U}(\mathbf{x}_{2,m})$ consists of a series of (multi-)controlled Z operations. Combined with the initial state $|\psi_0\rangle=|+\rangle^{\otimes 2n}$, it is clear to see that our encoding is equivariant with respect to both symmetries. 
As for the equivariant ansatz, we select generators that commute with $\hat{U}_{Z_2}$ and $\hat{U}_\Phi$. These include local operators $\Big\{\sum_i \hat{Y}_i, \sum_i \hat{X}_i\hat{X}_{i+1}, \sum_i \hat{Y}_i\hat{Y}_{i+1}, \sum_i \hat{Z}_i\hat{Z}_{i+1}\Big\}$ as well as global operators $\Big\{\hat{X}^{\otimes 2n}, \hat{Z}^{\otimes 2n}, \prod_{i=1}^n \mathrm{SWAP}_{i,i+n}, \hat{H}^{\otimes 2n}, \ldots\Big\}$. We can extend this pool by noting that if operators $\hat{A}$ and $\hat{B}$ both commute with $\hat{C}$, then so does the product $\hat{A}\hat{B}$. This allows us to insert operators such as $\Big\{\prod_{i=1}^n \mathrm{SWAP}_{i,i+n}\cdot\hat{X}^{\otimes 2n}, \hat{H}^{\otimes 2n}\cdot\hat{Z}^{\otimes 2n}, \prod_{i=1}^n \mathrm{SWAP}_{i,i+n}\cdot\hat{H}^{\otimes 2n}, \ldots\Big\}$ into the pool. We select from this same pool of operators when choosing the invariant measurement observable.

We train the quantum model by minimizing the mean squared error (MSE) between the model predictions and the data labels over the training data set, $\mathcal{L}_{\mathrm{MSE}} = \frac{1}{M}\sum_{m=1}^M (ah_\theta(x_m)+b-y_m)^2$, where $M$ is the number of training samples and $a$ and $b$ are trainable scaling and bias parameters. 
%
%
In this variational GQML approach [denoted as $\text{QNN}_\mathcal{U}$ as in Fig.~\ref{fig:architectures}(a)], we optimize the parameters within the variational ansatz and measure a single observable.

Next, we present a second option for GQML, involving the measurement of multiple observables with adjustable weights [$\text{QNN}_\mathcal{M}$, Fig.~\ref{fig:architectures}(b)]. 
We take inspiration from the ideas presented by Molteni \textit{et al.} \cite{molteni2024exponential}. There, the authors prove quantum advantages for the task of learning quantum observables of the form $\hat{O}(\boldsymbol{\alpha}) = \sum_{k=1}^K \alpha_k\hat{P}_k$, where $\alpha_k$ are real, unknown coefficients and $\hat{P}_k$ are Pauli strings. 
Specifically, the task is to learn a model $h(x)$ which approximates some unknown concept $f^{\boldsymbol{\alpha}}(x)=\mathrm{tr}[\rho_H(x)\hat{O}(\boldsymbol{\alpha})]$. Here, $\rho_H(x) = \hat{U}(x)|0\rangle\langle0|\hat{U}^\dagger(x)$, with $\hat{U}(x)=e^{i\hat{H}(x)}$. This problem is provably hard for classical algorithms under certain conditions, but the proposed quantum protocol can learn the concept with a training set of polynomial size \cite{molteni2024exponential}. 

Let us detail the method, adapting it specifically to our task of determining whether pairs of barcodes are correlated or not. 
We first encode the barcodes into quantum states in the same way as the standard GQML approach; $|\phi_{x_m}\rangle = \hat{\mathcal{U}}_{\phi}(x_m)|\psi_0\rangle$. We then build a feature vector $\varphi(x_m)$ for each data point $x_m$ by measuring the expectation values of a set of observables $\hat{\mathcal{O}}$, directly on the quantum states, $\varphi(x_m) = [\langle \hat{O}_1\rangle_m,\langle \hat{O}_2\rangle_m,\ldots,\langle \hat{O}_K\rangle_m]$, where $\langle\hat{O_i}\rangle_m = \langle x_m|\hat{O}_i|x_m\rangle$. We select this set of observables from the aforementioned pool of equivariant operators, such that our model maintains symmetry invariance. 
We then define the model as $h(x_m)=\boldsymbol{\alpha} \cdot \phi(x_m)$, where $\boldsymbol{\alpha}$ is a vector of real-valued weights. We make use of the LASSO (least absolute shrinkage and selection operator) method, a machine learning model that finds the optimal weights to minimize the loss function as \cite{friedman2010regularization}
\begin{equation}
    \mathrm{LASSO}: ~ \min_{\boldsymbol{\alpha}}\frac{1}{2M}\sum_{m=1}^M \Big(\boldsymbol{\alpha}\cdot\phi(x_m)-y_m \Big)^2+\lambda||\boldsymbol{\alpha}||_{l_1}, 
\end{equation}
where $\lambda$ is a regularization constant which constrains the size of the weights. In this case, our labels $y_m=f^\alpha(x_m)$ represent samples from our concept class. By optimizing the loss function, we can determine the combination of observables that lead to a clear separation between correlated and non-correlated samples $x_m = (\mathbf{x}_{1,m},\mathbf{x}_{2,m})$.

We proceed to compare the performance of the two different quantum architectures, $\text{QNN}_\mathcal{U}$ and $\text{QNN}_\mathcal{M}$. For the variational GQML, we selected the generators $\Big\{\sum_i \hat{Y}_i, \sum_i \hat{X}_i\hat{X}_{i+1}, \sum_i \hat{Y}_i\hat{Y}_{i+1}, \prod_{i=1}^n \mathrm{SWAP}_{i,i+n}\Big\}$ to build our model [Fig.~\ref{fig:architectures}(a)], balancing layers of trainable local operations with the measurement of a non-local SWAP operator in an attempt to capture global correlations between the input pair. The variational parameters were optimized using the Adam optimizer with a learning rate of 0.1, and training was performed over 200 epochs. As for the adaptive measurement protocol, we selected a set of $K=10$ measurement observables from our equivariant pool of operators, including most of the operators already listed, and the weights of the LASSO model were optimized via coordinate descent. For both architectures, we used a training dataset of $10$ samples from each class $(M=20)$, and tested on $40$ samples from each class, thus studying the models' ability to generalize to unseen data from relatively few samples. We trialled the two quantum models on pairs of $4\times 4$ barcodes, averaging results for each model over $10$ trials (for each trial, variational parameters/weights were randomly initialized and a different training set was selected). The results are shown in Fig.~\ref{fig:measVunitary}. 
We can see that both models successfully train; the training loss of $\text{QNN}_\mathcal{U}$ steadily decreases over the training period, while the $\text{QNN}_\mathcal{M}$ model converges extremely quickly. However, we observe a stark difference in generalization; while the adaptive measurement model achieves virtually perfect test accuracy across all runs, the variational circuit model struggles to reach test accuracies over 60\%. $\text{QNN}_\mathcal{U}$ may be overfitting the training data, indicating that the features encoding the correlations are difficult to extract using the parameterized quantum circuit framework. However, measuring a linear combination of specific observables proves to be much more successful in distinguishing correlated and uncorrelated samples. The struggles of $\text{QNN}_\mathcal{U}$ are even more pertinent as we increase the system size, with an increasing number of ansatz layers required just to train the model, with similarly poor generalization. Based on these trials, we proceed with the adaptive measurement approach as the GQML protocol to test on larger images.


\textit{The results.---}We analyze the described GQML workflows and compare them to classical similarity testing with Siamese neural networks \cite{Bromley1993,Koch2015SiameseNN}. To emphasize the resilience of the separation between quantum and classical learners, we present generalization over the number of training samples and system size.
\begin{figure}[t!]
    \includegraphics[width=\linewidth]{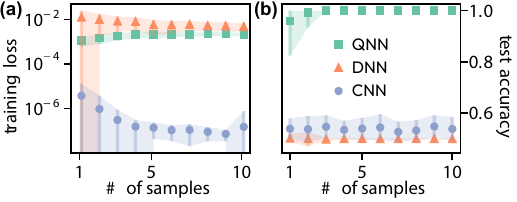}
    \caption{Generalization of quantum and classical learners for the barcode similarity problem shown as a function of the number of samples from each class used in training. We show final training loss \textbf{(a)} and test accuracy \textbf{(b)} of quantum and classical networks, trained and tested on pairs of 1024-pixel barcodes (20-qubit system). Classical benchmarks correspond to deep (DNN) and convolutional (CNN) Siamese neural networks, which learn the similarity via a metric distance evaluated in feature space (see details in the main text). 
    }
    \label{fig:generalization}
\end{figure}
In Fig.~\ref{fig:generalization}, we show the training loss and test accuracy of the $\text{QNN}_\mathcal{M}$ against those for classical networks, classifying pairs of $32\times 32$ barcodes. We consider two types of networks with Siamese structures, namely a deep neural network (DNN) and a convolutional neural network (CNN). Siamese neural networks consist of two identical sub--networks that share the same weights and architecture. Let $f_{\theta}(\mathbf{x})$ represent the function learned by the sub--network parametrized by weights $\theta$. Given a dataset pair $x_m = (\mathbf{x}_{1,m},\mathbf{x}_{2,m})$, the network computes the outputs $\mathbf{h}_{1,m} = f_\theta(\mathbf{x}_{1, m})$ and $\mathbf{h}_{2,m} = f_\theta(\mathbf{x}_{2, m})$. The idea is to map similar inputs close to each other in feature space while mapping dissimilar inputs far apart. To achieve this, we compute the commonly used Euclidean distance metric between the two outputs, $d(\mathbf{h}_{1,m}, \mathbf{h}_{2,m}) = ||\mathbf{h}_{1,m} - \mathbf{h}_{2,m}||^2$. The learning objective is to minimize the loss function $\mathcal{L}_{\mathrm{MSE}} = (1/M)\sum_{m=1}^M (\Tilde{y}_m-y_m)^2$ where $\Tilde{y}_m$ is the predicted label for the pair $(\mathbf{x}_{1,m},\mathbf{x}_{2,m})$. Note that the mean square error is chosen for better comparison with the quantum models. We used the following training and test parameters for the simulations with quantum and classical networks. We average over 50 randomizations of the training dataset for each training data set size, training over (up to) 300 epochs for both models (noting again that far fewer epochs were required for the convergence of $\text{QNN}_\mathcal{M}$). For each run, we randomize the $M$ samples $x_m$ selected for the training dataset, with $\frac{M}{2} = 1, 2, \ldots, 10$ selected from each category (correlated and uncorrelated samples). Testing is done on 40 unseen samples from each class. The error bar represents the standard deviation over the training randomizations. Note that the ``tearing'' effect of the lower error bars in Fig.~\ref{fig:generalization}(a) is due to the semi--logarithmic scale.

Both quantum and classical ML protocols showcase good training performances across all datasets; all three models reach 100\% training accuracy at all dataset sizes $M$, with MSE for CNN converging to particularly low values. The real differences between the architectures are seen in their generalization, where both classical models struggle to perform on unseen data, achieving test accuracies only slightly better than random guessing. This suggests that the classical models are prone to overfitting the training data, failing to discover the true features underlying the correlations between the barcodes. Several adjustments were attempted to improve the classical performance; we tested different loss functions (binary cross--entropy), added dropouts to the classical neural networks, used cosine similarity as a metric, and varied the batch size. None of these settings significantly improved results. Meanwhile, the quantum model again generalizes excellently at this larger system size; training on just 3 samples from each class $(M=6)$ is enough to classify all the pairs of 1024-pixel barcodes loaded into $2n=20$-qubit register.
\begin{figure}[hb]
    \includegraphics[width=\linewidth]{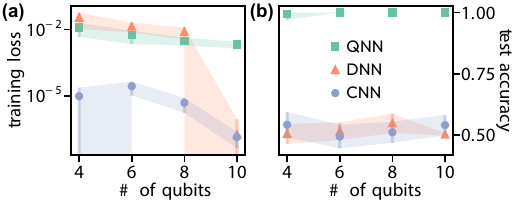}
    \caption{Generalization of learners for the barcode similarity problem shown as a function of system size (number of qubits $n$ for each image, $2n$ in total). Both, training loss \textbf{(a)} and test accuracy \textbf{(b)} remain steady across all system sizes, indicating that the quantum--classical separation is ubiquitous for smaller and larger pairs of barcodes. The plots show simulations run with the same parameters as described in Fig.~\ref{fig:generalization}, with 5 training samples taken from each class ($M=10$).}
    \label{fig:scaling}
\end{figure}

Next, we compare the ability of different models to detect image correlation at different system sizes fixing the number of training samples. The results are shown in Fig.~\ref{fig:scaling}. We plot the average training loss and test accuracy as a function of the number of qubits $n$ used to encode each barcode within the quantum model; an $N$-pixel barcode requires $n=\log_2 N$ qubits. Both quantum and classical models show no significant performance variation across the range of image dimensions, showcasing that quantum-classical separation persists at different system sizes. While we fix the number of training samples to 5 from each class ($M=10$), all other parameters and the running procedure are the same as the ones of Fig.~\ref{fig:generalization}.


\textit{The explanation.---}Previously we have shown that the problem of barcode similarity testing can be solved with excellent generalization. This is ensured by the considered dataset, symmetries, and trained quantum models. Let us unravel why GQML did so well when learning the similarity (global correlations), and show that the task can be related to problems leading to a proven superpolynomial speedup. To form the dataset we employed the strategy presented by Raz and Tal \cite{RazTal2019}, motivated by Aaronson \cite{Aaronson2010}, and developed to show a separation between $\BQP$ and $\PH$. Specifically, we take a normal multivariate probability distribution $\tilde{\mathcal{P}}_{\mathrm{A}}(x) \defeq \mathcal{N}(0, \varepsilon \cdot \bm{\Sigma})$ over $\tilde{x} \in \mathbb{R}^{N} \times \mathbb{R}^{N}$ with zero mean, variance of $\varepsilon$, and covariances defined as a $2 \times 2$-block  matrix $\bm{\Sigma} = [\mathds{1}^{\otimes n}, \hat{H}^{\otimes n}; \hat{H}^{\otimes n}, \mathds{1}^{\otimes n}]$, for $n = \lceil \log_2 N\rceil$. Drawing samples from this distribution $\tilde{X} \sim \tilde{\mathcal{P}}_{\mathrm{A}}(\tilde{x})$ we get a list of correlated vectors. This can be seen as drawing the $N$-dimensional first part of the sample $\tilde{\mathbf{x}}_1$ and the second part being related via the Hadamard transform as $\tilde{\mathbf{x}}_2 \approx \hat{H}^{\otimes n} \tilde{\mathbf{x}}_1$. Following Ref.~\cite{RazTal2019} we perform truncation of vector entries and get the samples $x_{|y=0}$. Next, sampling the distribution $\tilde{\mathcal{P}}_{\mathrm{B}}(x) \defeq \mathcal{N}(0, \varepsilon \cdot [\mathds{1}^{\otimes n}, \bm{0}; \bm{0}, \mathds{1}^{\otimes n}]$), then performing the same truncation gives us uncorrelated samples from effectively independent distributions, $x_{|y=1}$. Raz and Tal then showed that following the quantum algorithm for the forrelation problem by Aaronson and Ambainis \cite{AaronsonAmbainis2015} one can solve this Promise-$\BQP$ problem of distinguishing forrelated and uniform distributions, calling a single query to load the data as a quantum state [similar to our feature map $\hat{\mathcal{U}}_{\phi}(x)$]. This $O(1)$ query complexity for distinguishing distributions showcases advantage over $\tilde{\Omega}(\sqrt{N})$ for the classical query complexity \cite{AaronsonAmbainis2015,bravyi2021classical}. The essence of the quantum algorithm is the estimation of the overlap between quantum representations of $\mathbf{x}_1$ and $\mathbf{x}_2$ as
\begin{equation}
\label{eq:forrelation}
F = |\langle \phi_{\mathbf{x}_1} | \hat{H}^{\otimes n} |\phi_{\mathbf{x}_2} \rangle|^2 = |\langle 0|\hat{H}^{\otimes n} \hat{U}_{\phi}^\dagger(\mathbf{x}_1)  \hat{H}^{\otimes n} \hat{U}_{\phi}(\mathbf{x}_2) \hat{H}^{\otimes n} |0 \rangle|^2,
\end{equation}
which takes high values for pairs from $P_{\mathrm{A}}(x)$ as we are effectively undoing the Hadamard transform on $\mathbf{x}_2$. Conversely, for uncorrelated samples $\hat{H}^{\otimes n}$ simply reshuffles the amplitudes, and $F$ remains exponentially small in $N$. This represents a feature that can be used by the GQML classifier. To pick up this feature of forrelated distributions during similarity testing QNNs require learning a specific observable for constructing a decision boundary, corresponding to $\hat{\mathcal{O}}(\boldsymbol{\alpha}^*) = \hat{H}^{\otimes 2n} \cdot \left(\prod_{i=1}^n \mathrm{SWAP}_{i,i+n}\right)$ applied on $|\phi_{\mathbf{x}_1} \rangle \otimes |\phi_{\mathbf{x}_2} \rangle$. Such a Hermitian operator is readily available in $\text{QNN}_\mathcal{M}$ model, and can be learnt by LASSO optimization. At the same time, we note that there is no equivariant unitary transformation and observable such that $\hat{W}^\dagger(\theta)\hat{\mathcal{O}}\hat{W}(\theta)$ represents the required product of SWAPs and Hadamards. This explains why $\text{QNN}_\mathcal{U}$ model does not generalize well for the task.

We stress that forrelation also underpins the first example of provably efficient quantum machine learning solvers of $\BQP$ problems in the kernel-based operation mode \cite{Jager2023}. Jäger and Krems have shown that by using a feature map specific to $k$-forrelation one can produce a classifier that outperforms any classical analog. In our work, we show that GQML achieves a generalization advantage without tailored feature maps, and aided by simple symmetries can distinguish distributions and patterns of global correlations.


\textit{Further discussion.---}So, answering the question in the title, can we indeed get an advantage in learning global correlations for pixelated images or barcodes? The answer is still no, at this point. The reason lies in the classical nature of the data that we are dealing with. For pixels that are stored in a memory we can apply the same steps as in the quantum algorithm and obtain an efficient quantum-inspired solution as the Hadamard transform is easy to apply on a classical machine. The drawback of non-oracular data loading is that the process is much easier to dequantize as long as we look inside a black box \cite{stoudenmire2023grovers}. It is also known that the 2-fold forrelation is classically simulable \cite{bravyi2021classical}. However, there is a scenario where a similar approach can give a genuine advantage. When generalizing Eq.~\eqref{eq:forrelation} to 4-fold forrelation, the problem becomes classically intractable \cite{bravyi2021classical}. The data loading becomes more complicated, as consecutive applications of Hadamards and multi-CZ gates moves the embedding outside of the REW state subspace. At the same time, it hints to us that there are patterns (i.e. more complicated barcodes) and distributions that can be loaded into quantum states, and classified by building a decision boundary based on generalized forrelation.




\textit{Conclusion.---}In this paper, we described a geometric quantum machine learning (GQML) approach that is able to learn an optimal algorithm for similarity testing between two vectors (visualized as images or barcodes). Our approach embedded symmetries to enable generalization from just a few samples. Unlike existing GQML algorithms, we focused on symmetry-aware measurement adaptation, which we demonstrated to outperform unitary adaptation of observables. We compared the performance of our GQML approach to that of classical deep neural networks employing a Siamese architecture. Our findings suggest that quantum machine learning surpasses classical deep learning, particularly in scenarios involving correlated distributions and limited training data. This conclusion was supported by the established maximal separations between the $\BQP$ complexity class and the polynomial hierarchy for classifying samples from forrelated and uniform distributions. Our work suggests the area where quantum machine learning can excel and gain advantage over classical machine learning.

\textit{Acknowledgement.---}We acknowledge the funding from UK EPSRC award under the Agreement EP/Y005090/1.




%

\end{document}